\documentclass[conference]{IEEEtran}

\usepackage{cite}
\usepackage[dvipsnames]{xcolor}
\usepackage{graphicx}
\usepackage{svg}
\usepackage{array}
\usepackage{caption}
\usepackage[utf8x]{inputenc}
\usepackage[normalem]{ulem}
\usepackage{multirow}

\begin{document}
\title{Performance Analysis of CP2K Code for Ab Initio Molecular Dynamics}

\author{Dewi Yokelson, Nikolay Tkachenko, Robert Robey, Ying Wai Li, Pavel A. Dub}
\author{\IEEEauthorblockN{Dewi Yokelson,
Nikolay V. Tkachenko, Robert Robey, Ying Wai Li and Pavel A. Dub}
\IEEEauthorblockA{Los Alamos National Laboratory\\
Los Alamos, New Mexico\\
Email: \{dyokelson,ntkachenko,brobey,yingwaili,pdub\}@lanl.gov\\}}

\maketitle

\begin{abstract}
Using a realistic molecular catalyst system, we conduct scaling studies of ab initio molecular dynamics simulations using the CP2K code on both Intel Xeon CPU and NVIDIA V100 GPU architectures. We explore using process placement and affinity to gain additional performance improvements. We also use statistical methods to understand performance changes in spite of the variability in runtime for each molecular dynamics timestep. We found ideal conditions for CPU runs included at least four MPI ranks per node, bound evenly across each socket, and fully utilizing processing cores with one OpenMP thread per core, no benefit was shown from reserving cores for the system. The CPU-only simulations scaled at 70\% or more of the ideal scaling up to 10 compute nodes, after which the returns began to diminish more quickly. Simulations on a single 40-core node with two NVIDIA V100 GPUs for acceleration achieved over 3.7x speedup compared to the fastest single 36-core node CPU-only version, and showed 13\% speedup over the fastest time we achieved across five CPU-only nodes.
\end{abstract}


\IEEEpeerreviewmaketitle

\section{Introduction}
Ab initio molecular dynamics (MD) is one of the most efficient methods to study chemical reactions in solution and solid phases in an accurate and unbiased manner~\cite{Kirchner2012}. In this method, atomic forces are computed on-the-fly from electronic structure calculations and then used to determine the microscopic trajectory of each individual atom through integration of Newton’s equations of motion.  Over the past few decades, various quantum chemistry packages such as Gaussian~\cite{g16}, VASP~\cite{VASP1, VASP2}, Q-CHEM~\cite{QCHEM}, Quantum Espresso~\cite{QE}, CP2K ~\cite{doi:10.1063/5.0007045}, BigDFT~\cite{BIGDFT}, NWChem~\cite{NWChem}, and TeraChem~\cite{TerraChem} have been developed to run ab initio molecular dynamics on both Central Processing Units (CPUs) and Graphics Processing Units (GPUs). Due to the exponential growth of the size of the wave function with the number of quantum particles and quality of the description of the system, ab initio molecular dynamics simulations on CPUs and GPUs are memory-intensive and are challenging to parallelize. Effective scalability is limited to tens of compute nodes containing CPUs and/or GPUs for most popular commercial codes. The number of performance parameters that can be used, such as number of nodes, MPI ranks, OpenMP threads, and number of GPU units, presents a bewildering array of options to scientific users for gaining the best performance with each application and chemical system to be studied. Complicating the run-time options are the large memory requirements for some calculations which add even more constraints. In addition, techniques such as process placement and vectorization which can yield better performance are typically unknown and unused by most scientists. Here, we investigate the performance of one of the most popular codes to simulate homogeneously-catalyzed reactions – CP2K~\cite{doi:10.1063/5.0007045, cat1, cat2, cat3, cat4, JPCA}.

CP2K is an open-source software package written in Fortran for electronic structure and molecular dynamics calculation. The code has excellent performance for atomistic simulations of liquids, solid-states, and other molecular and biological systems. One of the main features of the code is the Quickstep module~\cite{VANDEVONDELE2005103}, allowing the use of both atom-centered Gaussian orbitals and plane waves as a basis for the solution of the Schr\"odinger equation. This feature helps when analyzing large systems such as liquid solutions or chemical reactions with the explicit inclusion of solvent molecules. In this work, we analyze the multi-level parallel performance of the CP2K code on CPUs and GPUs for a scientific study of the intermediate complex found in the catalytic Milstein ester hydrogenation reaction in methanol~\cite{Gusev2020RevisedMO}.

Recent peer-reviewed works by Kühne, et al.~\cite{doi:10.1063/5.0007045} and VandeVondele, et al.~\cite{VANDEVONDELE2005103} include a good, basic performance analysis of CP2K. Both of these are heavily focused on comparing solution time between different solver methods. However, there is a demand for more computer architecture performance-focused studies centered around optimization. We see in both~\cite{Reid2013OptimisingCF,Bethune2014CP2KPF} that there is a need to accelerate specific CP2K implementations even further. This can be achieved by focusing on the computing side - running more in depth performance analysis and optimization.

Our primary goal for this work is to understand and optimize the performance of CP2K on the current high-performance computing processors with large processor counts per node, wider vector units, and GPUs. Demonstrating good scalability of the CP2K application will support its use on larger computing systems, enabling faster, more complex calculations. We will look at whether process placement and affinity can squeeze out more performance. The trade-off between using OpenMP threads and MPI ranks will also be explored in conjunction with process placement. Then we will investigate the benefits of wider vector units with the latest compilers and processors.

The effects of optimizations on CPUs can be difficult to ascertain. The timing results for applications on CPUs can have significant variation from run to run that can obscure small improvements in performance. And yet many small improvements can become significant and can be important for extending scaling of performance to larger processor counts. We apply the Student's T-test~\cite{StudentBiometrika} to analyze the timing from a small set of runs to get a confidence level for the impact of a change on performance. We use a black-box approach where instrumentation is not added to the application code.

We also seek to understand and reduce the variability in measurement of run performance so that we can get either a higher confidence in the results or be able to use a smaller number of runs for the same confidence level.  Our first approach to reducing the timing variability will be using a consistent process placement for the timing runs versus the default placement by the operating system kernel. We use the F-test~\cite{snedecor1989statistical} to compare the variance of two distributions to determine whether there is a significant reduction in the variation of the timing data due to process placement. Note that we use multiple molecular dynamics timesteps in a single run for some of the statistics in our analysis. This avoids having to make multiple runs of the code, but it does add another source of variability. As an alternative to run-time, we also look at other measurable statistics such as memory bandwidth of the application.

\section{Background}

This section contains more information on the type of scientific calculations being done in this particular CP2K problem. It also discusses the type of calculations in the chemical system being studied and the performance expectations of the calculations. 

\subsection{CP2K in a Molecular Catalytic System}

\begin{figure}
    \centering
    \includegraphics[scale=0.5]{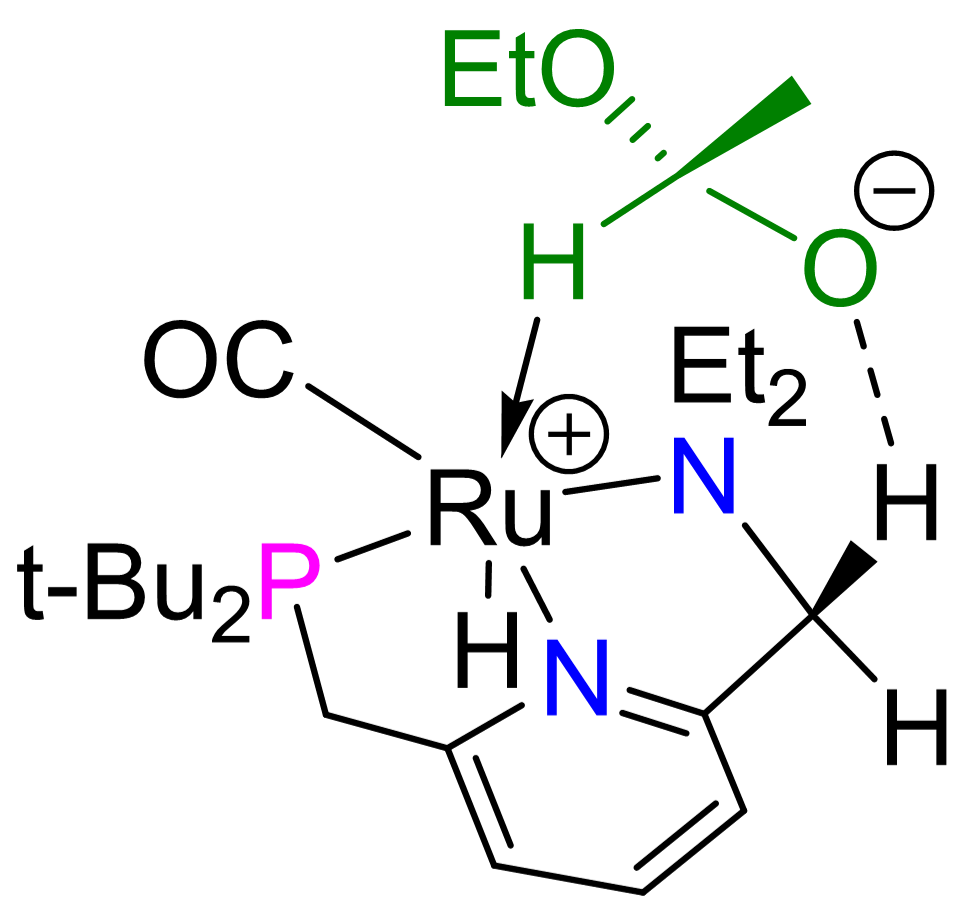}
    \caption{Intermediate in the Milstein ester hydrogenation reaction~\cite{Gusev2020RevisedMO}}
    \label{fig:intermediate}
\end{figure}

For the performance tests, we studied the common intermediate in the Milstein ester hydrogenation reaction in methanol, Fig.~\ref{fig:intermediate} \cite{Gusev2020RevisedMO}. The intermediate complex was surrounded by 130 molecules of methanol solvent. The total number of atoms in the simulation box was 856 including one \textit{d}-element. Simulation boxes were created using the Packmol package~\cite{https://doi.org/10.1002/jcc.21224} with 2.0 Å tolerance parameter and 23 Å × 23 Å × 23 Å box size. The periodic boundary conditions were applied to all $x-$, $y-$ and $z-$directions to mitigate the surface effects. Ab initio calculations of electronic structures were carried out using density functional theory (DFT), in which energies and forces were evaluated with the Generalized Gradient Approximation (GGA) BLYP~\cite{PhysRevA.38.3098,PhysRevB.37.785} functional corrected by the D3~\cite{doi:10.1063/1.3382344} dispersion term. Goedecker–Teter–Hutter (GTH) pseudopotentials~\cite{PhysRevB.54.1703, PhysRevB.58.3641} as well as DZVP-MOLOPT-SR-GTH (Ru atom) and DZVP-GTH (all other atoms) mixed Gaussian and plane wave basis sets~\cite{doi:10.1063/1.2770708} were used. The plane wave cutoff was set to 700 Ry, while Gaussians were projected with the plane wave cutoff of 140 Ry. Then the isothermal-isobaric ensemble (NPT) was sampled in the molecular dynamics simulations of the system. The Nose–Hover thermostat~\cite{PhysRevA.31.1695, doi:10.1063/1.447334} was employed at a temperature of 373 K, while the barostat pressure was set to 50 bar.

\subsection{Performance Implications}

Ab initio molecular dynamics codes are computationally heavy due to the need to converge the wave function in the ab initio DFT calculation within each MD timestep. The intensive parts of the calculations are the matrix diagonalization that typically scales as the cubed of the number of electrons of the system; it also involves calculating FFTs which are both challenging operations on high-performance computing systems. The performance limitations come from many factors. For example, the convergence of the wave function is an iterative procedure where the runtime depends on the quality of the initial trial wave function; high intensity floating point operations in some parts of the calculations such as the matrix operations, low arithmetic intensity kernels but heavy communications in the molecular dynamics parts, and the choice of density functionals that determines the accuracy of the calculations. 
Not all parts of the calculation can be parallelized and the remaining serial sections, according to Amdahl's Law~\cite{amdahl1967validity}, will often limit the overall scalability. The need for storing the wave function of quantum systems makes ab initio molecular dynamics memory intensive, which adds complications to performance optimization of the code. However, they parallelize well during certain compute-intensive phases. The usual approach in most ab initio molecular dynamics codes is to parallelize the key computationally heavy kernels. 

\section{Methodology}

This section describes the performance experiments that were conducted. We discuss the hardware used as well as both the CPU and GPU trials measuring runtime and memory bandwidth. We discuss the three different approaches to extracting results from the timing data and the pros and cons of each. 

\subsection{Hardware Specifications}

\begin{table*}
\caption{CPU and GPU architectures used in this study.}
\centering
\begin{tabular}{ c | c | c | c | c | c | c | c } 
\hline
\textbf{CPU Architecture} & \textbf{Processor} & \textbf{Clock (GHz)} & \textbf{Sockets} & \textbf{Cores / Socket} & \textbf{Year} & \textbf{\# GPUs} & \textbf{GPU Architecture}\\ 
\hline
\multirow{2}{*}{x86\_64} & Broadwell & \multirow{2}{*}{2.1} & \multirow{2}{*}{2} & \multirow{2}{*}{18} & \multirow{2}{*}{2016} & \multirow{2}{*}{0} & \multirow{2}{*}{N/A}\\
& Intel Xeon E5-2695 v4  &&&&&\\
\hline
\multirow{2}{*}{x86\_64} & Broadwell & \multirow{2}{*}{2.2} & \multirow{2}{*}{2} & \multirow{2}{*}{12} & \multirow{2}{*}{2016} & \multirow{2}{*}{0} & \multirow{2}{*}{N/A}\\
& Intel Xeon E5-2650 v4 &&&&&\\
\hline
\multirow{2}{*}{x86\_64} & Haswell  & \multirow{2}{*}{2.6} & \multirow{2}{*}{2} & \multirow{2}{*}{10} & \multirow{2}{*}{2014} & \multirow{2}{*}{1} & \multirow{2}{*}{NVIDIA Tesla V100 PCIE 32 GB}\\
& Intel Xeon E5-2660V3 &&&&&\\
\hline
\multirow{2}{*}{x86\_64} & Cascade Lake  & \multirow{2}{*}{2.5} & \multirow{2}{*}{2} & \multirow{2}{*}{20} & \multirow{2}{*}{2019} & \multirow{2}{*}{2} & \multirow{2}{*}{NVIDIA Tesla V100 PCIE 32 GB}\\
& Intel Xeon Gold 6248 &&&&&
\label{table:hardware}
\end{tabular}
\end{table*}

The affinity experiments were run on a cluster containing Intel Xeon Broadwell processors. Each node contains two sockets and a corresponding Non-Uniform Memory Access (NUMA) domain. Each socket has 18 physical cores and with hyperthreading turned off, the virtual processing core count is also 18 cores. Single node GPU trials were run on two different systems. The first is an Intel Xeon Haswell CPU with one NVIDIA Tesla V100 GPU. The other is an Intel Xeon Cascade Lake Gold CPU that has two NVIDIA Tesla V100 GPUs. See Table~\ref{table:hardware} for a more detailed specification of the CPU and GPU architectures. 

The memory bandwidth experiments were run on a different cluster of Intel Xeon Broadwell processors. These similarly contain two sockets with a corresponding NUMA region. However, each of these sockets contains 12 cores with hyperthreading turned off as well. We also ran these experiments on the Cascade Lake GPU system described above. See Table~\ref{table:hardware} again for these specifications.

On all hardware, CP2K was compiled with gcc, gfortran, make, and cmake. Supporting libraries that were installed include: BLAS, LAPACK, MPI, SCALAPACK, FFTW, LIBINT, libsmm, libxsmm, CUDA, libxc, ELPA, PEXSI, QUIP, PLUMED, spglib, SIRIUS, FPGA, COSMA, and LibVori. Install instructions on the CP2K GitHub repository provide more information on how to configure each of these libraries~\cite{cp2kbuild}. We ran only the psmp version which is the hybrid OpenMP/MPI version without debug settings.

\subsection{Affinity}

Affinity is an important consideration when optimizing the performance of an application. Affinity goes beyond determining the best combination of the number of ranks and threads for the architecture. Affinity, also known as pinning or binding, is the scheduling of individual or groups of processes or threads onto specific hardware~\cite{Robey}. This section describes how we changed the numbers of threads and ranks, but also experimented with different pinning techniques. 

\subsubsection{Determining the Rank and Thread Count}
Some initial optimization and scaling work was completed on this ab initio molecular dynamics CP2K problem prior to this project. This yielded ideal performance of 2 MPI ranks per node and 16 OpenMP threads. However, because the particular cluster we used had nodes with 18 processors per socket, we changed the OpenMP thread count accordingly. As we scaled MPI ranks up, we adjusted the OpenMP threads to be fewer per rank, so there would be one thread per processor. We determined the numbers based on the hardware specifications and techniques in~\cite{Robey}. See Table \ref{table:experiments} for a more detailed specification of the parameters used. 

\subsubsection{Process Placement}
Oversubscription of processes on compute processors is a common issue when ranks are not explicitly forced into certain locations. This can cause performance lags as you may have five MPI ranks on one node or socket and none on another, even when you specify to use a certain number of compute nodes. If no placement is specified, MPI's default behavior is to let the operating system determine the process placement, as well as move the rank around during runtime. We refer to this unassigned behavior as \textit{default} as the only directions specified are the number of ranks per node.

One aspect of process placement is forcing a certain number of ranks on each compute resource. In our \texttt{mpirun} command we used the \texttt{--cores-per-socket} flag to specify this. For example, in a run on 10 nodes, with 20 ranks, we used \texttt{--cores-per-socket=1} so that one MPI rank would be bound to one socket, two per node. 

Another aspect of process placement is the order that ranks are distributed to each node. We analyze two different types of process distribution in our experiments. The first is \textit{round robin} distribution, where successive ranks are assigned to successive nodes. The second is \textit{block} distribution where successive MPI ranks are placed on the same node, so in essence they are grouped together in order. There are potential performance implications of each of these distributions depending on the application as ranks might be more likely to operate on the same memory or not, therefore the physical location could make a difference. 

We analyzed the process placement with the StreamTriad process placement tool from~\cite{Robey}. In this way, we discovered that by default the Slurm workload manager assigned the MPI ranks in a round robin fashion, with successive ranks on different nodes. Adding \texttt{--distribution=block} forced the successive MPI ranks to be assigned to nodes in order. This means that if there are two ranks per node, 0 and 1 are on the same node. 

\begin{table}[h!tb]
\caption{Some of the process placement and affinity settings used on the 36-core Intel Broadwell nodes.}
\centering
\begin{tabular}{c | c | c | c} 
\hline
\textbf{Nodes} & \textbf{Ranks} & \textbf{OpenMP Threads} & \textbf{Distribution}\\ 
\hline
1 & 2 & 18 & default, round robin, block\\
\hline
5 & 10 & 18 & default, round robin, block\\
\hline
10 & 20 & 18 & default, round robin, block\\
\hline
10 & 40 & 9 & default, round robin, block\\
\hline
10 & 60 & 6 & default, round robin, block\\
\hline
10 & 180 & 2 & round robin \\
\hline
10 & 180 & 1 & round robin \\
\hline
25 & 50 & 18 & default, round robin, block
\label{table:experiments}
\end{tabular} 
\end{table}

\subsection{Extracting Results from Noisy Timing Data}

\begin{figure}
\includegraphics[width=\linewidth]{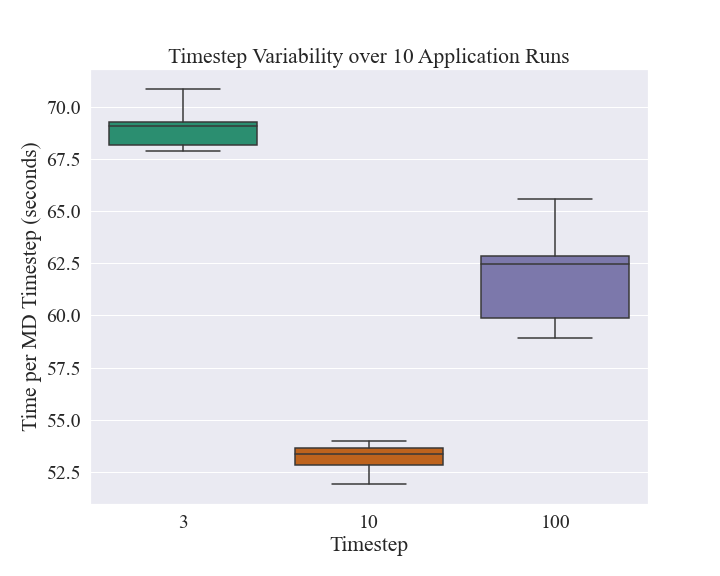}
\caption{Variability in runtime at selected MD timesteps, each averaged over 10 application runs. The variation in the mean values of around 25\% is due to the difference in convergence time required in the DFT calculations at each timestep, while the variation for the runtime for the 100\textsuperscript{th} timestep varies about 12\%. To identify the effectiveness of our optimization approaches we are looking to detect a consistent change in performance that is a fraction of either of these variations, requiring a more rigorous statistical approach.}
\label{fig:timesteps}
\end{figure}

While conducting these tests we found that each MD timestep varied to a rather large extent, with a standard deviation approaching 10\% of the average in many cases. Thus we considered three approaches for measuring the runtime of CP2K in this context. The first was the total runtime of the application, summing all MD timesteps and having a single number. However, due to the high variability we observe, this requires running the application many times per configuration in order to get enough data points to ensure there is a difference in the performance with that configuration. The second was comparing each MD timestep with itself across application runs, i.e. looking at how timestep 3 performed across multiple application runs. Fig.~\ref{fig:timesteps} shows how this data can vary for sample timesteps 3, 10, and 100, though not to as large a degree as when we look at all of the timesteps in a single run. This method still requires numerous application runs, which can be time consuming. Thus we settled on the third approach, this was taking the average of a small sample of timesteps in a single application run, but applying statistical techniques to ensure this sample would yield accurate results.

\begin{figure}
\includegraphics[width=\linewidth]{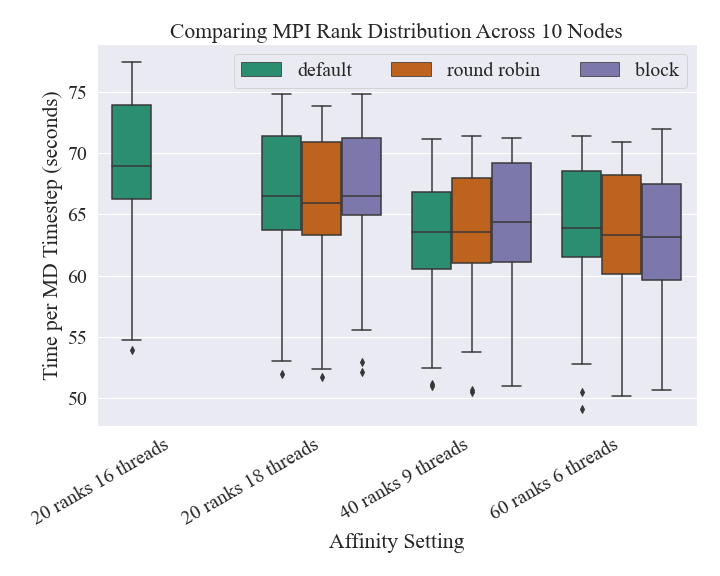}
\caption{Variability in runtime over 35 MD timesteps in a single simulation using different affinity settings. The plot shows the distribution of data points in quartiles, with outliers indicated by separate points (diamonds).}
\label{fig:distribution}
\end{figure}

Runtime averages and confidence intervals were calculated with a sample size of 35 MD timesteps. This did not include timesteps 0 and 1 which are initialization steps and thus, outliers. 30 is considered the proper sample size to guarantee a normal distribution and provide accuracy when used in statistical methods. We found that with such high variability and small sample size we needed a method to be able to confidently say whether an array of MD timesteps was significantly faster or not. Fig.~\ref{fig:distribution} shows the variability per MD timestep. 

\subsubsection{T-tests and F-tests}

Our solution for determining whether a change had a significant effect on the average MD timestep runtime was to perform a T-test. A T-test is a statistical routine in which a p-value is calculated. The p-value represents the probability that two arrays of numbers are from a different distribution or not. We chose this because it is recommended for use with a small sample size. A T-test uses a t-statistic and sample means and variances instead of a z-statistic and population means and variances. In our case, we would calculate a p-value from 35 MD timesteps, while a typical production run might normally consist of 20,000 MD timesteps or more.

If the calculated p-value is below a certain significance level, typically $\alpha = 0.05$, then the two means are considered significantly different. In other words, the two underlying populations from which these samples were taken likely have different means. For example, if we sample 35 MD timesteps from a run with certain affinity settings and 35 from another, we can compare these two with a T-test and determine the probability that there is a significant difference between the two populations.

An F-test to compare two variances is used to determine whether the variance in one array of numbers is significantly different than the variance in another. From this, we find the probability that two variances are significantly different. This is useful in extracting our results as well because the more consistent we know the measurements to be, the fewer data points are required to be confident in our results. Strictly speaking, an F-test assumes a normal distribution and might not work as well for other statistical distributions. But for our purposes, the F-test is sufficient to get an initial assessment whether there is a difference in the variance. If we wanted a higher quality result, more data points would be more important than using a more sophisticated variance test method.

\subsection{Performance Measurements on GPUs}

Based on the nature of this problem and promising indicators from K{\"u}hne et al.~\cite{doi:10.1063/5.0007045}, we believed that running this application on a GPU could provide substantial performance benefits. In addition to being able to achieve significant speedup we were interested in comparing the variability of the runtime per molecular dynamics timestep with that of the CPU-only build. This may help us to better understand if our performance is improved based on changes we make to the configuration. We ran trials on two types of compute nodes: a single node with a CPU connected to a GPU, another is a single node with a CPU connected to two GPUs (see Table \ref{table:hardware}). This included using a single MPI rank and scaling the OpenMP threads as well as scaling MPI ranks and threads. 

\subsection{Measuring Memory Bandwidth}

We also looked at the memory bandwidth and total runtime as measured by the LIKWID tool~\cite{Treibig2010LIKWIDLP}. We chose LIKWID because it is lightweight --- with minimal overhead or slowdown --- and because of its mpi-run wrapper which can measure the memory bandwidth of hybrid multi-threaded and MPI applications. LIKWID is also includes support for NVIDIA GPU architectures which makes it especially useful in comparing our CPU and GPU results. 

Memory bandwidth can still vary from run-to-run due to the variation in the placement of the memory in the heap. But the variability of memory bandwidth should in general be less than the variability of run time measurements for an application. Since the performance of applications are generally memory bandwidth limited, the change in memory bandwidth can serve as an indicator that a change has made a positive effect or not. While an improvement in memory bandwidth should provide the same improvement in run time for many code sections, it is not clear that this can be assumed for overall application performance with a complex mixture of kernels such as the CP2K application.

\section{Results}

In this section we summarize the results of our experiments, including discussion about the speedup we achieved on the CPU-only version, the significant speedup on the GPU version, and comparing the different approaches to measurement including runtime and memory bandwidth.

\begin{figure}
    \centering
    \includegraphics[width=\linewidth]{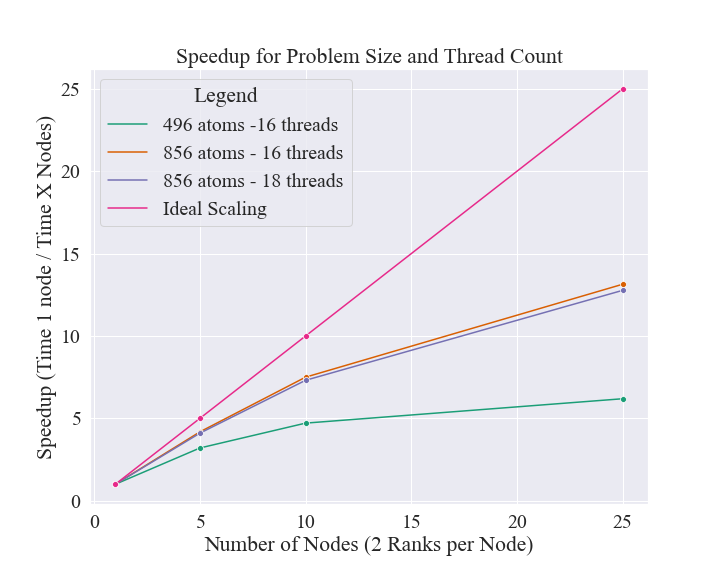}
    \caption{Scaling behavior of two problem sizes. There is better scaling efficiency on the larger problem size. The speedup increases more slowly for the smaller problem size.}
    \label{fig:scaling}
\end{figure}

\subsection{CPU Results}

Better affinity settings alone led to about 9.5\% speedup on the average MD timestep on the CPU. This was achieved by binding MPI ranks to sockets and ensuring there were OpenMP threads to utilize each core on the socket. Even without affinity, just increasing the OpenMP thread count from 16 to 18 rendered about 4\% speedup as this kept all 18 processors on the socket busy. We were able to compare these times and achieve a p-value of $1.37 \times 10^{-20}$, or a 99.999...\% probability that the small change alone had a significant impact on the runtime. This coincides with results from~\cite{Reid2013OptimisingCF} depicting peak performance of their benchmark problem when all processors on the node are utilized.  

Comparing the different rank distribution techniques was less straightforward. We found that round robin distribution almost always outperformed both default and block. P-values were smaller when comparing either distribution setting to the default - which implies higher confidence levels that including those settings helped shorten the runtime. Looking at round robin vs block distribution, we found that round robin usually outperformed block. It was only for 40 ranks, 9 threads that the T-test yielded a p-value below the significance value of 0.05, and this was 0.006 or a 99.4\% probability that they are from different distributions.  

In addition, the observed standard deviation between the timesteps decreased by over five percent with better affinity settings. This was not as much of a reduction we were hoping for though. The F-test comparing the 18 thread version and 16 thread version returned a probability of only 20\% that the variances are significantly different. These mixed results are the reason driving us towards the memory bandwidth and GPU measurements. We hoped to be able to achieve less variability.

Fig.~\ref{fig:scaling} depicts the scaling of one of the best affinity settings --- 18 OpenMP threads, two MPI ranks per node, one rank bound to each socket --- and compares it with the performance of 16 threads with the same total number of ranks but no affinity settings. We compared with the 16 threads case as it had previously been found to be the best settings for this project before our study. The best affinity settings, while achieving speedup, scaled at a similar rate to when using only 16 threads and no binding. 

Of note is also the smaller problem size, which was also run prior to this study, with 496 atoms, 16 threads and no affinity settings. We never see as high a rate of speedup, and it begins to slow down much sooner as we increase the nodes and ranks. Similar scaling results were obtained by K{\"u}hne et al.~\cite{doi:10.1063/5.0007045}, where the authors showed that the deviation from the ideal scaling is less for larger system. We conclude that we are running out of work to keep all the processors busy for this problem.

\subsection{GPU Results}

\begin{figure}
    \centering
    \includegraphics[width=\linewidth]{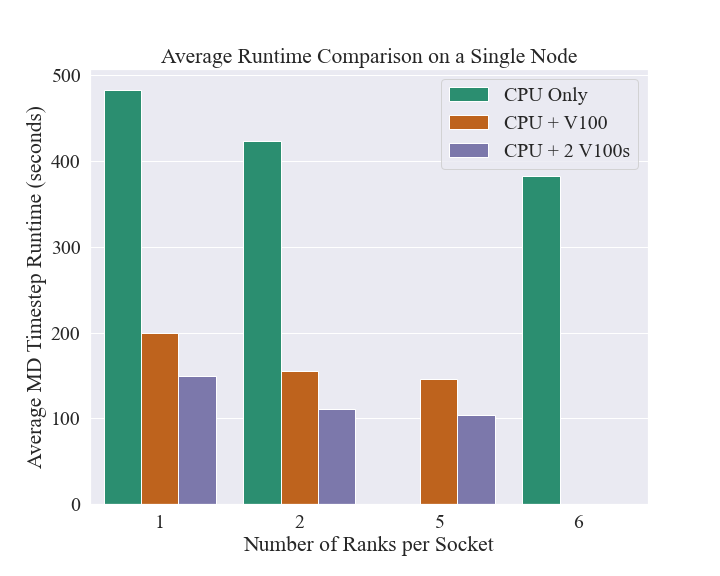}
    \caption{Comparison of CPU and GPU average MD timestep runtimes. The x-axis shows how many ranks were bound to each socket, i.e. 5 ranks per socket means 10 total MPI ranks were used. All these trials used one OpenMP thread per core. We compare 5 and 6 ranks per socket so as to assign thread count evenly to cores as these systems have different numbers of cores.}
    \label{fig:gpucpu}
\end{figure}

GPU performance on the Cascade Lake/V100 node was significantly faster than a single CPU node. When comparing the fastest single-node runs, using 12 MPI ranks and 3 OpenMP threads on the GPU node we achieved over 3.7x speedup compared to running 10 MPI ranks and 4 OpenMP threads on the Broadwell CPU node. The runtime of each MD timestep on the GPU node was roughly equivalent to 5 CPU-node performance. This was achieved by scaling both MPI ranks and threads. Fig.~\ref{fig:gpucpu} depicts the average runtime for a MD timestep on a single Intel Xeon Broadwell CPU node compared with the single Intel Xeon Haswell node containing one NVIDIA V100 and the single Intel Cascade Lake node containing two NVIDIA V100s. These results align with the potential for 2-4x speedup reported by K{\"u}hne et al.~\cite{doi:10.1063/5.0007045} for their GPU implementation.

\begin{figure}
    \centering
    \includegraphics[width=\linewidth]{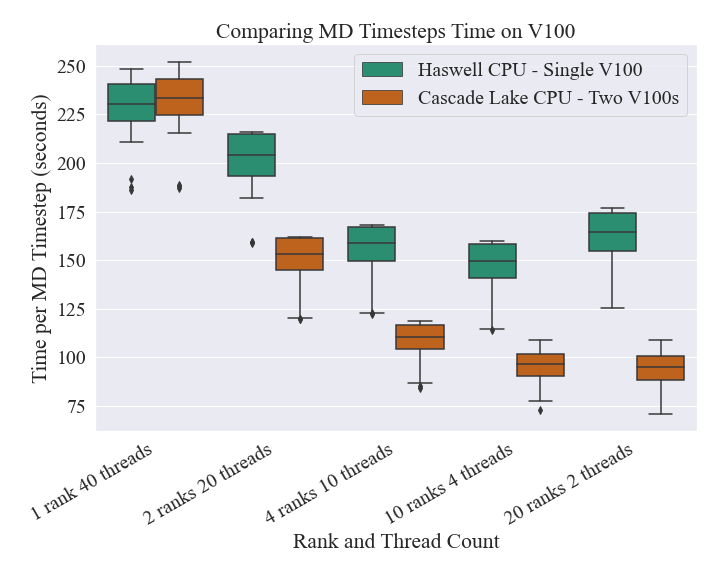}
    \caption{MD timestep runtimes in seconds on two different nodes: (green) a node with one Intel Haswell CPU connected with one NVIDIA V100 GPU, and (brown) a node with one Intel Cascade Lake CPU connected with two NVIDIA V100 GPUs.}
    \label{fig:gpuvar}
\end{figure}

Variability of runtime on the GPU was somewhat reduced over the CPU - where the CPU had standard deviation around 8-10\% of the mean, GPU trials typically yielded a standard deviation closer to 7-9\% of the mean runtime. See Fig.~\ref{fig:gpuvar} for the runtime variability results with varying rank and thread counts. As an example, conducting an F-test comparing the fastest GPU run with the most similar CPU performance (5 nodes, 10 ranks, 18 threads) yielded a probability of 0.055\% that the variance is the same. Therefore the variability was significantly reduced running on GPU architecture.

\begin{figure}[h]
    \centering
    \includegraphics[width=\linewidth]{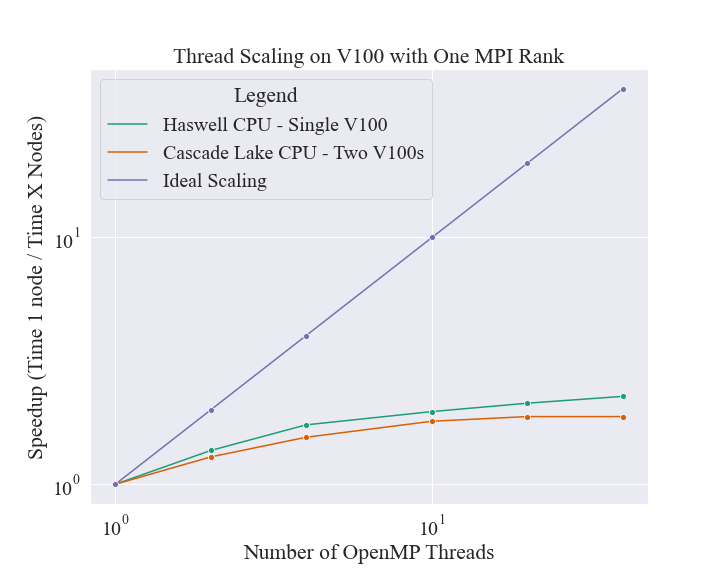}
    \caption{The thread scaling behavior on a single Intel Xeon Haswell CPU node with two NVIDIA V100 GPUs shows significantly diminishing returns when using more than 4 OpenMP threads.}
    \label{fig:gputhreadscaling}
\end{figure}

Varying the number of OpenMP threads on the GPU node resulted in much less speedup than with MPI ranks, even at a higher number of OpenMP threads. Fig.~\ref{fig:gputhreadscaling} shows the speedup when running only one MPI rank but scaling the OpenMP thread count. Scaling the threads only did not yield significant performance benefits, especially after 4 threads. The Cascade Lake-dual V100s system outperformed the Haswell-single V100 system based on runtime up through 10 threads, with the T-test returning p-values below the significance value. Once we run 20 threads, the Haswell-single V100 system performs better with a mean timestep of 228.09 compared to the mean of 231.25 and a p-value of $5.5 \times 10^{-5}$. Although the Cascade Lake-dual V100s system was generally faster, it always performed slightly worse in terms of speedup gained with increased thread count.

\subsection{Memory Bandwidth Results}

Memory bandwidth proved to be the most easily understood measure, as each trial run provides a single measurement. After examining the variability across multiple application runs on a single 24-core Broadwell CPU-only node, the standard deviation was observed to be less than 1\% of the mean. With a smaller standard deviation we can be more confident that a new measurement provides assurance that a significant change was made. However, while higher memory bandwidth implies faster runtime, it is not a guarantee when it comes to small differences. Table~\ref{table:cpumemory} shows that this is not always the case, and that more parameters must be considered. The best bandwidth measurement and runtime are highlighted in bold for the two settings we investigated, but they are not always from the same application run.

Table~\ref{table:gpumemory} lists some of the results of running on the Cascade Lake / V100 node. These experiments produced similar variance numbers as on the CPU. We can see from the results how the much higher bandwidth on the GPU nodes can have an effect on the performance. With the highest bandwidth value measured for 20 ranks, 2 threads (utilizing all cores on the node) on the GPU outperforming the fastest CPU-only version of 12 ranks, 2 threads (again, utilizing all cores on this node) by 2x. Despite only doubling the memory bandwidth, the total runtime speedup was 3.4x when comparing the fastest runs of these same two rank/thread settings.

\begin{table}[h!tb]
\caption{Sample Memory Bandwidth Results on a 24-core Broadwell CPU node.}
\centering
\begin{tabular}{ c | c | c } 
\hline
 \textbf{Setting} &\textbf{Bandwidth (Mbytes/s)} & \textbf{Total Runtime (s)} \\ 
\hline
\multirow{3}{*}{12 ranks 2 threads} & 15961.8202 & 16464.42 \\
 & \textbf{16228.4178} & 16450.08 \\
 & 15978.3218 & \textbf{16201.49}\\
\hline
\multirow{3}{*}{6 ranks 4 threads} & \textbf{8092.9993} & \textbf{28883.01} \\
 & 8054.4494 & 29045.81 \\
 & 8043.7417 & 29035.68 
\label{table:cpumemory}
\end{tabular} 
\end{table}

\begin{table}[h!tb]
\caption{Sample Memory Bandwidth Results on a 40-core Cascade Lake / V100 node.}
\centering
\begin{tabular}{ c | c | c } 
\hline
 \textbf{Setting} &\textbf{Bandwidth (Mbytes/s)} & \textbf{Total Runtime (s)} \\ 
\hline
\multirow{3}{*}{20 ranks 2 threads} & \textbf{33295.9147} & 4789.012\\
 & 32551.9212 & \textbf{4760.265} \\
 & 33278.5851 & 4788.942\\
\hline
\multirow{3}{*}{10 ranks 4 threads} & 29161.4547 & \textbf{4783.195} \\
 & \textbf{29395.8641} & 4859.71 \\
 & 29143.5516 & 4785.793 
\label{table:gpumemory}
\end{tabular} 
\end{table}

\section{Conclusion}

Getting application scaling on high-processor count CPUs or on multiple GPUs is important but can be challenging for scientific productivity. We review our approach to improving application performance; developing a process to extract reliable information from noisy performance data and reducing their variability.

Our primary goal was to see if we could improve the performance of the CP2K application on current high-end multi-processor CPUs and on GPUs. We demonstrated that better affinity settings on the CPU were able to improve performance by nearly 10\%. We also showed that reserving an extra processor for the system did not help either performance or reduce the variability in the run-times --- by using all the processors, we were able to get an additional 4\% speedup. Our recommendation for the most efficiency on CPU-only nodes is to run on 5-10 nodes, with at least four MPI ranks per node, bound evenly across the sockets of the node. Then calculate the OpenMP threads so that there is one per core on the node. These performance improvements will likely be similar on the many other simulation problems run with CP2K, and may be extensible to other similar chemistry applications.

The GPU results were even more promising, showing significant speedup of about 3.7x compared to CPU-only simulations. We found the performance of having two GPUs per node on the newer Intel CPU hardware to exceed that of a single GPU per node on older hardware. Even though the use of multiple MPI ranks helped us achieve this speedup, there are still a lot of possible variations in how to run with multiple GPUs in combination with MPI ranks and OpenMP threads. We have observed lower run-time variability of applications on GPUs. But what about multi-GPU runs where the CPU and network variability is also important? Knowing whether increasing GPUs per node or increasing the number of nodes with GPUs can provide better returns would be very useful. In the future we would also like to explore the use of CUDA Multi-Process Service to better utilize the GPU compute capacity with cooperative MPI jobs.

Another of our goals was to extract more confident results as to whether changes were beneficial despite the noisy performance data. We used long-standing statistical methods to compare the mean and the variance of the noisy data to get a higher confidence that a change gave improvement and by how much. Optimizing performance of an application is often the accumulation of many small improvements rather than due to just a single change. Evaluating the benefit of each small change can be challenging when the run-to-run variation in performance is larger than that of the change.

The last of our goals was to understand and possibly reduce the variability in performance measurements. Besides the speedup from affinity settings, we anticipated that pinning processes to a location would also reduce the run-time performance variability. Indeed, we saw a greater than 5\% reduction in the variance. This could be because there was only a small amount of process movement by the operating system during the short run-time of the test cases and that longer runs might see more process movement, leading to more run-time variability.

Since most application kernels are bandwidth limited, we selected memory bandwidth as an alternate performance measurement. While memory bandwidth is the most appropriate surrogate measurement for performance, there are many other measurements from hardware counters that could yield insights into whether there are performance improvements and what might be the best targets for further optimizations.

Another interesting avenue of exploration would be to both recompile CP2K and its dependent libraries with more aggressive vectorization flags on different compilers and see if vectorization can accelerate the code further. The recent availability of 512-bit wide vector units, twice the width of prior vector units, might improve the performance of some computational kernels. Beyond the vectorization compiler flags we could use profiling tools such as Intel Advisor to find problematic loops and ensure all possible loops are being vectorized.

\section{Acknowledgement}
We thank the Los Alamos National Laboratory (LANL) Institutional Computing program for access to HPC resources. Dewi Yokelson is a recipient of a 2021 LANL Parallel Computing Summer Research Internship and performed this work as part of a summer internship. Ying Wai Li is supported by LANL's Laboratory Directed Research and Development Program under project number 20210087DR. LANL is operated by Triad National Security, LLC, for the National Nuclear Security Administration of U.S. Department of Energy (Contract No. 89233218CNA000001).
\bibliographystyle{IEEEtran}
\bibliography{refs}

\end{document}